\documentclass[manuscript]{aastex}
\usepackage{times}
\usepackage{amssymb}
\usepackage{epsfig}
\usepackage{rotating}
\usepackage{lscape}

\def\ga{\mathrel{\raise0.35ex\hbox{$\scriptstyle >$}\kern-0.6em
\lower0.40ex\hbox{{$\scriptstyle \sim$}}}}
\def\la{\mathrel{\raise0.35ex\hbox{$\scriptstyle <$}\kern-0.6em
\lower0.40ex\hbox{{$\scriptstyle \sim$}}}}
\def\co{CO~{\it J}=$1-0$ }

\def\cosix{CO~{\it J}=6-5 }

\def\loj{low-{\it J}~}

\newcommand{\hi}{H{\sc i}}
\newcommand{\kms}{km~s$^{-1}$}

\shorttitle{Molecular gas in $z > 6.5$ Lyman-$\alpha$ emitters}
\shortauthors{Wagg, Kanekar, \& Carilli}
\slugcomment{draft version: April 7, 2009}

\begin{document}

\title{The molecular gas content of $z> 6.5$ Lyman-$\alpha$ emitters}

\author{
Jeff~Wagg$^{1}$, Nissim~Kanekar$^{1}$ and Christopher~L.~Carilli}

\affil{National Radio Astronomy Observatory, PO Box O, Socorro,
  NM 87801, USA; $^1$Max-Planck/NRAO Fellow}\email{jwagg@nrao.edu}

\begin{abstract}
We present results from a sensitive search for \co\ line emission in
two  $z> 6.5$ Lyman-$\alpha$ emitters (LAEs) with the Green Bank 
Telescope. \co\ emission was not detected from either object. For HCM~6A, 
at $z \sim 6.56$, the lensing magnification factor of $\sim 4.5$ 
implies that the CO non-detection yields stringent constraints on the 
\co\ line luminosity and molecular gas mass of the LAE, $L^\prime_{\rm CO} < 
6.1 \times 10^{9} \times (\Delta V/300)^{1/2}$~K~\kms~pc$^2$ and
$M_{\rm H_2} < 4.9 \times 10^{9} \times (\Delta V/300)^{1/2} \times (X_{\rm CO}/0.8) 
\: M_\odot$.
These are the strongest limits obtained so far for a $z \gtrsim 6$ galaxy. For IOK-1, 
the constraints are somewhat less sensitive, $L^\prime_{\rm CO} < 2.3 \times 10^{10} 
\times (\Delta V/300)^{1/2}$~K~\kms~pc$^2$ and $M_{\rm H_2} < 1.9
\times 10^{10} \times (\Delta V/300)^{1/2} \times (X_{\rm CO}/0.8) \: M_\odot$.
The non-detection of \co\ emission in HCM~6A, whose high estimated star formation
rate, dust 
extinction, and lensing magnification make it one of the best high-$z$ LAEs 
for such a search, implies that typical $z \gtrsim 6$ LAEs are likely to 
have significantly lower \co\ line luminosities than massive sub-mm galaxies 
and hyperluminous infrared quasars at similar redshifts, due to either a 
significantly lower molecular gas content or a higher CO-to-H$_2$ conversion factor.
\end{abstract}


\section{Introduction}

In recent years, multi-wavelength selection techniques have been successful 
in identifying different populations of star-forming galaxies at high redshifts, 
$z \gtrsim 6$ [see \citet{ellis08} for a recent review]. An important class 
of such galaxies consists of the Lyman-$\alpha$ emitters (hereafter LAEs), 
objects identified through their excess emission in narrow-band images centred 
on the redshifted Lyman-$\alpha$ wavelength
(e.g. \citealp{hu98,rhoads00,taniguchi05}). Follow-up Lyman-$\alpha$ 
spectroscopy has yielded accurate redshifts for a significant fraction 
of the LAE population (e.g. \citealp{taniguchi05,kashikawa06}), unlike most other 
high-$z$ star-forming galaxies (e.g. Lyman-break systems, sub-mm galaxies),
which typically only have photometric redshifts. In fact, an LAE at $z = 6.96$ 
\citep{iye06} has the highest confirmed spectroscopic redshift of all 
presently-known galaxies.

LAEs constitute a significant fraction of the star-forming galaxy population 
at $z \sim 6$, sufficient to reionize the Universe at earlier epochs (e.g. \citealp{fan06}).
The number density of LAEs and the typical shape of the Lyman-$\alpha$ 
emission line therefore provide important probes of physical conditions in the Universe 
around the epoch of reionization (e.g. \citealp{haiman99,haiman02,kashikawa06}).
Equally important, the steep drop in the space density of quasars at $z \gtrsim 6$ 
implies that they would be unable to produce the ultraviolet (UV) background 
radiation required to reionize the Universe (e.g. \citealp{fan01}). Star-forming 
galaxies like the LAEs are thus most likely to have been responsible for reionization 
(e.g. \citealp{yan04}). Understanding the factors that influence the
star-formation activity in these systems (e.g. the molecular gas content, 
the star formation efficiency, etc) is hence of much importance.

Recent studies of individual LAEs at $4.2 < z < 6.6$ have shown 
that these galaxies have a wide range of properties, with best-fit stellar masses 
ranging from $\sim 10^8 -10^{10} \: M_\odot$, ages from $\sim 3 - 1000$~Myr and 
metallicities from $0.005 - 1 \: Z_\odot$ (e.g. \citealp{chary05,lai07,finkelstein09a}). 
Evidence has also been found for significant amounts of dust in LAEs, with UV 
extinctions of $A_{\rm UV} \sim 0.5 - 5$~mag. (e.g. \citealp{chary05,finkelstein09a}). 
\citet{chary05} suggest that the observed decrease in the the cosmic star-formation 
rate at $z > 6$ (e.g. \citealp{stanway04}) might be explained by the presence of 
dust in the $z > 6$ LAE population. Typical star-formation rates (SFRs) in LAEs, as 
inferred from their UV continua, are low, a few tens of solar masses per year (e.g. 
\citealp{taniguchi05}); these are $2-5$ times larger than the SFRs derived from 
the resonantly-scattered Lyman-$\alpha$ line. While this difference could arise
from dust obscuration effects, \hi\ absorption by the damping wing of the 
Gunn-Petersen trough is likely to also contribute towards reducing the 
strength of the Lyman-$\alpha$ line for LAEs at $z > 6.5$ \citep{haiman02}.
We note that the tentative detection of the H$\alpha$ line in a single 
$z \sim 6.56$ LAE \citep{chary05} yielded a significantly higher SFR estimate 
($140 \: M_\odot$~yr$^{-1}$) than that obtained even from the rest-frame 
UV continuum ($\sim 9\: M_\odot$~yr$^{-1}$); this emphasizes the possibility 
that the SFRs in other LAEs might have been under-estimated due to dust extinction.

The detectability of high-$z$ galaxies like the LAEs relies on their undergoing 
an elevated level of star-formation activity, which naturally requires fuel 
in the form of molecular gas. Such gas is most effectively studied through observations 
of redshifted CO emission lines (e.g. \citealp{solomon05}). The luminosity in the 
\loj\ CO lines can be used to estimate the total molecular gas mass fueling 
the star-formation activity, while the CO line widths provide a measure of the 
dynamical mass of the galaxy. Studies of molecular gas at high redshifts, $z > 4$,
have so far focused on the most massive, far-infrared-luminous systems, the sub-mm galaxies 
or quasars (e.g. \citealp{schinnerer08,walter03}), and no information is available in 
the literature on the molecular gas content of ``normal'' star-forming galaxies, 
such as the LAEs. 

In the present work, we address this outstanding problem by conducting a sensitive 
search for low-excitation CO line emission in two $z > 6.5$ LAEs with the 
Green Bank Telescope (GBT), which allows us to place strong constraints 
on their total molecular gas masses. Throughout this {\it Letter}, we adopt a cosmological 
model with $(\Omega_\Lambda, \Omega_m, h) = (0.73, 0.27, 0.71)$ \citep{spergel07}.

\begin{figure*}[t!]
\centering
\epsfig{file=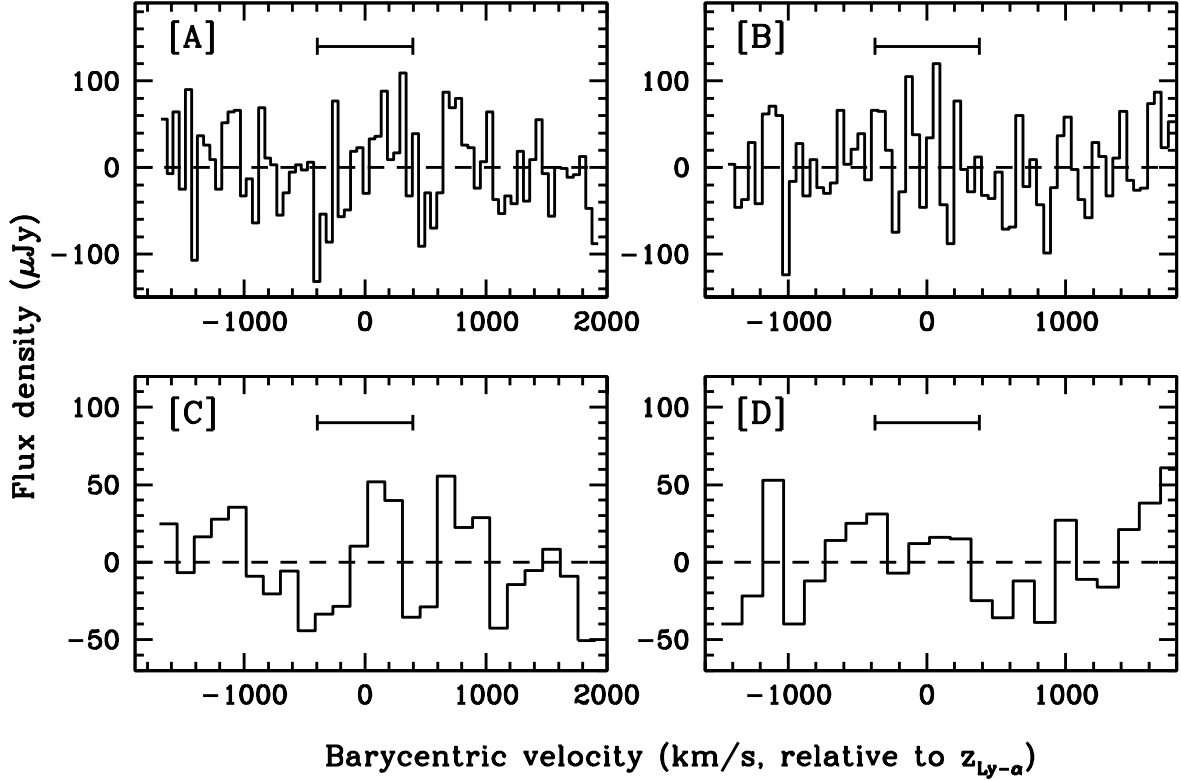,width=\textwidth}
\caption{GBT spectra in the redshifted \co\ line from HCM~6A (left panels, [A] 
and [C]) and IOK-1 (right panels, [B] and [D]). The spectra in the top and bottom panels 
have been smoothed to velocity resolutions of 50~km~s$^{-1}$ and 150~km~s$^{-1}$, respectively. 
The velocity scale of each spectrum is relative to the Ly$\alpha$ redshift, while
 the error bar in each panel indicates the redshift uncertainty (conservatively, 
$\Delta z = 0.01$).}
\label{fig:spectra}
\end{figure*}

\section{The targets}
\label{sec:sources}

Our two target LAEs were selected to have $z > 6.5$, to ensure that their redshifted 
\co\ lines could be observed with the sensitive GBT Ku-band receiver. 
The first system, HCM~6A at $z \sim 6.56$ \citep{hu02}, is perhaps the best candidate LAE 
for a search for molecular emission as it is strongly lensed by a foreground cluster 
(Abell~370, at $z \sim 0.375$; \citealp{kneib93}), with a magnification factor of 
$\sim 4.5$; this significantly improves the sensitivity to CO emission. It also 
has the highest estimated SFR of all known LAEs; \citet{chary05} found excess broad-band 
emission in a {\it Spitzer} IRAC~4.5$\mu$m image, compared to the continuum at 
other wavelengths, and argue that this is due to strong H$\alpha$ emission, with
an implied SFR of $140 \: M_\odot$~yr$^{-1}$. The SFR inferred from the UV 
continuum (uncorrected for dust effects) is significantly lower than this, 
$\sim 9 \: M_\odot$~yr$^{-1}$ \citep{hu02}, similar to values obtained in other 
LAEs at $z \sim 6.5$ (e.g. \citealp{taniguchi05}). If the SFR derived from the 
{\it Spitzer} image is correct, the discrepancy between the two SFR values implies 
a high dust extinction, $A_{\rm UV} \sim 2.6$~mag, consistent with the value 
independently derived from a fit to the broad-band photometric data (\citealp{chary05}; 
see also \citealp{schaerer05}). Such a high SFR would be similar 
to that of nearby ultra-luminous infrared galaxies (ULIRGs), which have typical 
molecular gas masses of $10^{9-10} \: M_{\odot}$ (e.g. \citealp{downes98}). 
Conversely, \citet{boone07} argue that the limits on the 850~$\mu$m and 
1.2~mm continuum flux densities of HCM~6A imply 
upper limits to the SFR in the range $10 - 90$~M$_\odot$~yr$^{-1}$, for 
assumed dust temperatures in the range $18-54$~K; these SFR estimates are 
somewhat lower than the values obtained by \citet{chary05} and \citet{schaerer05}.
 \citet{chary05} argue that the properties of HCM~6A are similar to those of 
other $z \gtrsim 6$ LAEs; the molecular gas properties derived for this 
object are thus likely to be representative of the entire population.

Our second target, IOK-1, was discovered in a Subaru survey for $z \sim 7$ 
LAEs \citep{iye06} and has the highest spectroscopically-confirmed redshift 
of all known galaxies. \citet{iye06} obtain a star-formation rate of 
$(10 \pm 2) \: M_{\odot}$~yr$^{-1}$ from the Lyman-$\alpha$ line luminosity. 
\citet{taniguchi05} find that the SFR estimated from the UV continua of 
$z \sim 6.5$ LAEs is typically $\sim 5$ times larger than that inferred
from the Lyman-$\alpha$ line, implying that the SFR of IOK-1 is likely to be 
$\gtrsim 50 \: M_\odot$~yr$^{-1}$, among the highest of the $z > 6$ LAE
population. This system has not so far been observed at sub-mm or infrared 
wavelengths and no information is thus available on its dust properties.

\section{Observations and data analysis}
\label{sec:obs}

The search for \co\ emission ($\nu_{\rm rest} = 115.2712$~GHz) from the two
LAEs was carried out with the GBT $12.0 - 15.4$~GHz Ku-band receiver 
between August and September 2008, during excellent summer observing conditions. 
The Auto-Correlation Spectrometer was used as the backend,
with two circular polarizations and a bandwidth of 200~MHz, sub-divided into 
8192 channels and centred on the redshifted \co\ line frequencies (14.48~GHz and
15.25~GHz). This yielded a total velocity coverage of $\sim 4000$~\kms\ and an 
initial velocity resolution of $\sim 1$~\kms, after Hanning-smoothing and re-sampling. 
Dual-beam nodding was used to calibrate the system bandpass, with a nodding timescale 
of 2~minutes. The telescope pointing was corrected every $2$~hours by observations 
of nearby bright calibrators. System temperatures were measured online by
firing a noise diode, and were typically found to lie in the range $22 - 30$~K. 
The standard flux calibrators, 3C48, 3C147 and/or 3C286 were used to verify 
the absolute flux scales at the observing frequencies; we estimate that the uncertainty 
in the flux calibration is less than 10\%. The total observing times were 
$\sim 25$~hours for  HCM~6A and $\sim 23$~hours for IOK-1, including overheads 
associated with dual-beam nodding, pointing and flux calibration. Note that the 
full-width-at-half-maximum of the GBT Ku-band beam is $\sim 50''$ (corresponding to 
a spatial scale of $\sim 280$~kpc at $z = 6.56$), implying that all \co\ emission 
from the LAEs (and any nearby galaxies at similar redshifts) would lie within 
the telescope beam. 

The data were analyzed using the {\sc GBTIDL}\footnote{http://gbtidl.sourceforge.net} 
data analysis package, following standard procedures. A second-order baseline was 
fit to each scan during the initial calibration, after which the data were examined 
on a scan-by-scan basis for residual baseline structure; this was done independently 
for both polarizations.  Conservatively, all scans showing baseline structure on 
scales of a few tens of MHz (that might mimic a putative spectral line) were edited 
out and not included in the final analysis; this resulted in the excision of 
$\sim 25$\% of data for each target. The final on-source integration times 
were $\sim 12$~hours for HCM~6A and $\sim 11$~hours for IOK-1. For each 
source, data from the two polarizations were combined together with appropriate 
weights, based on their root-mean-square (RMS) noise values to produce the final 
spectra. The spectra were then smoothed to coarser resolutions (50, 150 and 300~\kms)
to search for \co\ line emission.

\section{Results}
\label{sec:results}

\begin{table*}[t!]
\caption{Results from the GBT observations of HCM~6A and IOK-1.
\label{tab:tab1}}
\begin{center}
\begin{tabular}{ccccccc} 
\hline 
\hline 
Source & $z_{Ly\alpha}$ & $\nu_{{\rm CO} 1-0}$ & On-source time &
RMS$^\star$ &  $L^\prime_{\rm CO}$ ($3\sigma$) & $M_{\rm H_2}$ ($3\sigma$)  \\
       &                &  GHz                 &  Hrs.         &
$\mu$Jy &  $\times (\Delta V/300)^{1/2}$

& $\times (\Delta V/300)^{1/2} (X_{\rm CO}/0.8) \: M_\odot$ \\
       &                &                      &               &         &K~\kms~pc$^2$    & \\
\hline 
HCM~6A & 6.56 & 15.25 & 12 & 22.0 & $< 6.1 \times 10^9$~$^\dagger$     & $< 4.9 \times 10^{9}$~$^\dagger$ \\
IOK-1  & 6.96 & 14.48 & 11 & 17.3 & $< 2.3 \times 10^{10}$  & $< 1.9 \times 10^{10}$ \\
\hline
\end{tabular}
\vskip 0.1in
\noindent $^\star${}The quoted values of RMS noise, $L^\prime_{\rm CO}$ and $M_{\rm H_2}$ 
are at a velocity resolution of $300$~\kms. \\
\noindent $^\dagger${}For HCM~6A, the luminosity and mass limits have
been corrected for a magnification factor of 4.5. 
\end{center}
\end{table*}

The final spectra obtained towards HCM~6A and IOK-1 are shown in the four panels of 
Fig.~\ref{fig:spectra}, at velocity resolutions of $\sim 50$~\kms\ and $\sim 150$~\kms.
The final RMS noise values per $50$~\kms\ channel are 52~$\mu$Jy (HCM~6A) and 
$51$~$\mu$Jy (IOK-1), and per $150$~\kms\ channel are $31 \mu$Jy 
(HCM~6A) and $30 \mu$Jy (IOK-1). No evidence for \co\ line emission was seen
in the spectra at these (or other) velocity resolutions.

The non-detections of \co\ line emission place strong constraints on the CO line 
luminosities, which can, in turn, be used to derive limits on the total mass 
of cold molecular gas 
in the two LAEs. Following \citet{solomon05}, the \co\ line luminosity 
$L^\prime_{\rm CO}$ can be written as 
\begin{equation}
L^\prime_{\rm CO} = 3.25 \times 10^7 \: S_{\rm CO} \: \Delta V \: \nu^{-2}_{\rm obs}\:  D_L^2 \: (1+z)^{-3} \:\: ,
\end{equation}
where $\nu_{\rm obs}$ is the observing frequency (in GHz), $D_L$ is the 
luminosity distance (in Mpc), and $L^\prime_{\rm CO}$ is in K~\kms~pc$^{2}$. 
For our non-detections, $S_{\rm CO} \times \Delta V \equiv 3\sigma_{\Delta V} 
\times \Delta V$ (in Jy~\kms) gives the $3\sigma$ 
upper limit on the integrated flux density in the \co\ line, where $\sigma_{\Delta V}$ 
is the RMS noise at the velocity resolution $\Delta V$. We will assume a 
line width of $\Delta V = 300$~\kms, similar to the median observed width in 
high-$z$ quasars (e.g. \citealp{carilli06}). 
We then obtain $3\sigma$ \co\ line luminosity limits of $L^\prime_{\rm CO} 
< 6.1 \times 10^{9} \times (\Delta V/300)^{1/2}$~K~\kms~pc$^2$ (HCM~6A) 
and $L^\prime_{\rm CO} < 2.3 \times 10^{10} \times (\Delta V/300)^{1/2}$~K~\kms~pc$^2$ 
(IOK-1), after correcting the HCM~6A luminosity for the lensing magnification 
factor of 4.5 \citep{kneib93}.

To convert the \co\ line luminosity to an estimate of the total molecular gas 
mass requires the CO-to-H$_2$ conversion factor $X_{\rm CO}$ (e.g. \citealp{solomon05}). 
For virialized molecular clouds in a quiescent galaxy like the Milky Way, the 
conversion factor is typically $X_{\rm CO} \sim 4.6 \: M_{\odot}$~(K~\kms~pc$^2$)$^{-1}$ 
(e.g. \citealp{solomon91}); conversely, luminous infrared galaxies have far lower 
CO-to-H$_2$ conversion factors [$X_{\rm CO} \sim 0.8 \: M_{\odot}$~(K~\kms~pc$^2$)$^{-1}$; 
e.g. \citealp{downes98}].
For the nearby dwarf starburst galaxy, M82, \citet{weiss01} find that
the lowest values of the conversion factor are measured towards the central 
star-forming regions where the UV radiation field is most intense. 
Given that the two LAEs observed here appear to be undergoing elevated levels
of star-formation, we adopt the conversion factor $X_{\rm CO} = 
0.8 \: M_{\odot}$~(K~\kms~pc$^2$)$^{-1}$.
Note that this conversion factor is also used for CO line studies of 
high-$z$ sub-mm galaxies (e.g. \citealp{greve05}), allowing a direct 
comparison between the inferred molecular masses of sub-mm galaxies and LAEs.
Using this conversion factor then yields $M_{\rm H_2} < 4.9 \times 10^{9} \times 
(\Delta V/300)^{1/2}  \times (X_{\rm CO}/0.8) \: M_{\odot}$ for HCM~6A, and  
$M_{\rm H_2} < 1.9 \times 10^{10} \times (\Delta V/300)^{1/2} \times (X_{\rm CO}/0.8) 
\: M_{\odot}$ for IOK-1.  We emphasize that the upper limits to the molecular 
gas mass would be higher than the above values if the Milky Way conversion 
factor were applicable to these LAEs. The results of our observations
are summarized in Table~\ref{tab:tab1}.

Finally, the \co\ line luminosity and the FIR luminosity are correlated 
in nearby starburst and spiral galaxies \citep{gao04}, with 
${\rm log} L_{\rm FIR} = (1.26 \pm 0.08) \times {\rm log} L'_{\rm CO} - 0.81$ 
\citep{riechers06}. Our limits on the \co\ line luminosity can be combined with 
this relation to infer the FIR luminosity of the two LAEs and thence, their SFRs.
We obtain $L_{\rm FIR} < 3.3 \times 10^{11} \: L_\odot$ (HCM~6A) and 
$L_{\rm FIR} < 1.8 \times 10^{12} \: L_\odot$ (IOK-1), and SFRs 
of $< 66 \: M_\odot$~yr$^{-1}$ 
(HCM~6A) and $< 360 \: M_\odot$~yr$^{-1}$ (IOK-1), on using the relation
SFR~$ = 2 \times 10^{-10} \: (L_{\rm FIR}/L_\odot) \: M_\odot$~yr$^{-1}$ 
\citep{kennicutt98b}. The limit to the SFR in HCM~6A is similar to that 
obtained from the 1.2~mm and 850~$\mu$m continuum imaging 
(SFR~$< 10-90 \: M_\odot$~yr$^{-1}$; \citealp{boone07}). While 
our SFR limit in HCM~6A is significantly lower than the estimate of 
$140 \: M_\odot$~yr$^{-1}$ from the tentative detection of H$\alpha$ 
emission \citep{chary05}, it should be pointed out that our result depends 
on the assumption that the local correlation between \co\ line luminosity 
and FIR luminosity is applicable in $z > 6$ LAEs. Note, however, that some 
classes of high-redshift galaxies [e.g. the ``B$z$K'' galaxies, selected 
as outliers in plots of $({\rm B}-z)$ versus $(z-{\rm K})$ colors; 
\citealp{daddi04}] have larger CO line luminosities than predicted by 
the FIR-CO relation, with large molecular gas masses but low star formation 
efficiencies \citep{daddi08}. If the high-redshift LAEs are similar to 
the B$z$K galaxies, our limits to the \co\ line luminosities would imply 
even lower FIR luminosities, and star-formation rates, than those listed above.

\section{Discussion}
\label{sec:discuss}

These are the first constraints on the \co\ line luminosity and the 
molecular gas mass of $z \gtrsim 6$ LAEs, providing a new window into physical 
conditions in star-forming galaxies at the highest redshifts. The high inferred 
SFR and dust extinction in HCM~6A, as well as its large magnification factor, 
imply that it is one of the best candidates 
for a detection of CO emission in a high-$z$ LAE. The limit on its \co\ line 
luminosity obtained here is the deepest ever obtained for a $z \gtrsim 6$ 
galaxy. For the assumed CO-to-H$_2$ conversion factor, the molecular gas mass 
limit is within a factor of $\sim 2$ of the molecular gas mass of the Milky Way 
(e.g. \citealp{combes91}). While the limit on the \co\ line luminosity of IOK-1 
is not as strong, it is lower than the median luminosity in high-$z$ sub-mm 
galaxies ($L'_{\rm CO} = 3.8 \times 10^{10}$~K~\kms~pc$^2$; \citealp{greve05}) 
and similar to the luminosities of lower-redshift B$z$K galaxies (which have 
similar SFRs; e.g. \citealp{daddi08}). Our results for both LAEs rule out 
the presence of extreme \co\ line luminosities, such as those seen in sub-mm galaxies 
or hyperluminous IR quasars (e.g. \citealp{greve05,walter03,carilli07}).
LAEs thus appear to either contain significantly lower quantities of cold 
molecular gas or have significantly higher CO-to-H$_2$ conversion factors
than sub-mm galaxies or FIR-bright quasar hosts. The latter possibility cannot 
be ruled out as the CO-to-H$_2$ conversion factor is likely to depend on 
metallicity (e.g. \citealp{maloney88}), and quasar host galaxies and sub-mm 
galaxies appear to be dusty, metal-rich systems. 

Finally, it is clear that molecular gas must be present in LAEs to fuel the 
observed star-formation activity and dust reddening. In cases where the CO 
line emission is optically thick and thermalized, the flux density in the 
CO lines scales $\propto \nu^2 \propto J_{\rm U}^2$, where $J_{\rm U}$ is 
the rotational quantum number of the upper level. This suggests that, 
despite the sensitive limits obtained here, the $J_{\rm U} \ge 5$ CO lines 
may provide a more effective avenue to probe the molecular gas content of high-$z$ 
LAEs, with planned facilities like the Expanded Very Large Array (EVLA) and 
the Atacama Large Millimetre Array (ALMA). For example, for optically-thick,
thermalized emission, ALMA would be able to detect the \cosix\ line from 
a $z = 6.5$ star-forming galaxy with $M_{\rm H_2} = 3 \times 10^9 \: M_{\odot}$ 
in $3$~hours of on-source integration time. Unfortunately, the high kinetic
temperatures and densities required to raise the CO molecules to the high-$J$ 
excitation states are unlikely to be present in ``normal'' star-forming galaxies 
like the LAEs. For example, Fig.~9 of \citet{ao08} shows that the CO line 
intensities are sub-thermal at $J_{\rm U} \ge 5$ in almost all galaxies (including 
ULIRGs and sub-mm galaxies) with observations of these lines. The sole exception 
is the lensed quasar APM08279+5255, where a combination of AGN heating and very 
high gas densities appears to yield the high CO excitation \citep{weiss07}. 
This implies that it is likely to be difficult to detect the high-$J$ CO lines 
from $z \gtrsim 6$ LAEs even with ALMA. The $158 \mu$m fine-structure transition 
of ionized carbon may hence prove the best candidate for mapping the large-scale 
structure of high-$z$ star-forming galaxies and determining their dynamical masses 
\citep{walter08}. A search for this transition in HCM~6A is currently in progress
[Kanekar et al. (\textit{in prep.})].

\section{Summary}
\label{sec:sum}

We have carried out a deep GBT Ku-band search for \co\ line emission from two 
high-redshift Lyman-$\alpha$ emitters, HCM~6A at $z \sim 6.56$ and IOK-1 at 
$z \sim 6.96$. Our non-detection of CO emission from the lensed LAE, HCM~6A, 
implies strong constraints on its \co\ line luminosity and molecular gas mass, 
$L^\prime_{\rm CO} < 6.1 \times 10^{9} \times (\Delta V/300)^{1/2}$~K~\kms~pc$^2$ 
and $M_{\rm H_2} < 4.9 \times 10^{9} \times (\Delta V/300)^{1/2} \times 
(X_{\rm CO}/0.8) \: M_\odot$, the strongest obtained to date for a 
$z \gtrsim 6$ galaxy. In the case of IOK-1, the absence of a lensing 
magnification factor implies that the limits on the line luminosity and 
molecular gas mass are somewhat less sensitive than for HCM~6A; we obtain 
$L^\prime_{\rm CO} < 2.3 \times 10^{9} \times (\Delta V/300)^{1/2}$~K~\kms~pc$^2$ 
and $M_{\rm H_2} < 1.9 \times 10^{10} \times (\Delta V/300)^{1/2} \times 
(X_{\rm CO}/0.8) \: M_\odot$. If HCM~6A is a typical LAE at these redshifts, 
our results imply that LAEs are unlikely to show high \co\ line luminosities 
(such as those found in quasar host galaxies at similar redshifts), due to 
either a lower molecular gas content or a higher value of the CO-to-H$_2$ 
conversion factor.  It hence appears that observations of redshifted CO 
line emission are unlikely to be an effective means of studying gas dynamics 
in the less-luminous galaxies responsible for reionizing the Universe, even 
with upcoming arrays such as ALMA and the EVLA.

\section{Acknowledgments}

JW, NK and CC are grateful for support from the Max-Planck Society and the 
Alexander von Humboldt Foundation. We thank the Green Bank staff for the 
development and implementation of a new dynamic scheduling system which 
enabled us to obtain excellent high-frequency observing conditions for 
this project (GBT08B-043).  We thank the anonymous referee for
helpful suggestions on the original manuscript, and Frederic Boone for comments.   
The National Radio Astronomy Observatory is operated by 
Associated Universities, Inc, under cooperative agreement with the 
National Science Foundation. 
 
\bibliographystyle{apj}

\begin{thebibliography}{39}
\expandafter\ifx\csname natexlab\endcsname\relax\def\natexlab#1{#1}\fi

\bibitem[{{Ao} {et~al.}(2008){Ao}, {Wei{\ss}}, {Downes}, {Walter}, {Henkel}, \&
  {Menten}}]{ao08}
{Ao}, Y., {Wei{\ss}}, A., {Downes}, D., {Walter}, F., {Henkel}, C., \&
  {Menten}, K.~M. 2008, A\&A, 491, 747

\bibitem[{{Boone} {et~al.}(2007){Boone}, {Schaerer}, {Pell{\'o}}, {Combes}, \&
  {Egami}}]{boone07}
{Boone}, F., {Schaerer}, D., {Pell{\'o}}, R., {Combes}, F., \& {Egami}, E.
  2007, A\&A, 475, 513

\bibitem[{{Carilli} {et al.}(2007)}]{carilli07}
{Carilli}, C.~L. {et al.} 2007, ApJ, 666, L9

\bibitem[{{Carilli} \& {Wang}(2006)}]{carilli06}
{Carilli}, C.~L. \& {Wang}, R. 2006, AJ, 131, 2763

\bibitem[{{Chary} {et~al.}(2005){Chary}, {Stern}, \& {Eisenhardt}}]{chary05}
{Chary}, R.-R., {Stern}, D., \& {Eisenhardt}, P. 2005, ApJ, 635, L5

\bibitem[{{Combes}(1991)}]{combes91}
{Combes}, F. 1991, ARA\&A, 29, 195

\bibitem[{{Daddi} {et~al.}(2004){Daddi}, {Cimatti}, {Renzini}, {Fontana},
  {Mignoli}, {Pozzetti}, {Tozzi}, \& {Zamorani}}]{daddi04}
{Daddi}, E., {Cimatti}, A., {Renzini}, A., {Fontana}, A., {Mignoli}, M.,
  {Pozzetti}, L., {Tozzi}, P., \& {Zamorani}, G. 2004, ApJ, 617, 746

\bibitem[{{Daddi} {et~al.}(2008){Daddi}, {Dannerbauer}, {Elbaz}, {Dickinson},
  {Morrison}, {Stern}, \& {Ravindranath}}]{daddi08}
{Daddi}, E., {Dannerbauer}, H., {Elbaz}, D., {Dickinson}, M., {Morrison}, G.,
  {Stern}, D., \& {Ravindranath}, S. 2008, ApJ, 673, L21

\bibitem[{{Downes} \& {Solomon}(1998)}]{downes98}
{Downes}, D. \& {Solomon}, P.~M. 1998, ApJ, 507, 615

\bibitem[{{Ellis}(2008)}]{ellis08}
{Ellis}, R.~S. 2008, First Light in the Universe (Saas-Fee Advanced Course 36),
  259

\bibitem[{{Fan} {et~al.}(2006){Fan}, {Carilli}, \& {Keating}}]{fan06}
{Fan}, X., {Carilli}, C.~L., \& {Keating}, B. 2006, ARA\&A, 44, 415

\bibitem[{{Fan} {et al.}(2001)}]{fan01}
{Fan}, X. {et al.} 2001, AJ, 122, 2833

\bibitem[{{Finkelstein} {et~al.}(2009){Finkelstein}, {Rhoads}, {Malhotra}, \&
  {Grogin}}]{finkelstein09a}
{Finkelstein}, S.~L., {Rhoads}, J.~E., {Malhotra}, S., \& {Grogin}, N. 2009,
  ApJ, 691, 465

\bibitem[{{Gao} \& {Solomon}(2004)}]{gao04}
{Gao}, Y. \& {Solomon}, P.~M. 2004, ApJ, 606, 271

\bibitem[{{Greve} {et al.}(2005)}]{greve05}
{Greve}, T.~R. {et al.} 2005, MNRAS, 359, 1165

\bibitem[{{Haiman}(2002)}]{haiman02}
{Haiman}, Z. 2002, ApJ, 576, L1

\bibitem[{{Haiman} \& {Spaans}(1999)}]{haiman99}
{Haiman}, Z. \& {Spaans}, M. 1999, ApJ, 518, 138

\bibitem[{{Hu} {et~al.}(1998){Hu}, {Cowie}, \& {McMahon}}]{hu98}
{Hu}, E.~M., {Cowie}, L.~L., \& {McMahon}, R.~G. 1998, ApJ, 502, L99

\bibitem[{{Hu} {et~al.}(2002){Hu}, {Cowie}, {McMahon}, {Capak}, {Iwamuro},
  {Kneib}, {Maihara}, \& {Motohara}}]{hu02}
{Hu}, E.~M., {Cowie}, L.~L., {McMahon}, R.~G., {Capak}, P., {Iwamuro}, F.,
  {Kneib}, J.-P., {Maihara}, T., \& {Motohara}, K. 2002, ApJ, 568, L75

\bibitem[{{Iye} {et al.}(2006)}]{iye06}
{Iye}, M. {et al.} 2006, Nature, 443, 186

\bibitem[{{Kashikawa} {et al.}(2006)}]{kashikawa06}
{Kashikawa}, N. {et al.} 2006, ApJ, 648, 7

\bibitem[{{Kennicutt}(1998)}]{kennicutt98b}
{Kennicutt}, Jr., R.~C. 1998, ARA\&A, 36, 189

\bibitem[{{Kneib} {et~al.}(1993){Kneib}, {Mellier}, {Fort}, \&
  {Mathez}}]{kneib93}
{Kneib}, J.~P., {Mellier}, Y., {Fort}, B., \& {Mathez}, G. 1993, A\&A, 273, 367

\bibitem[{{Lai} {et~al.}(2007){Lai}, {Huang}, {Fazio}, {Cowie}, {Hu}, \&
  {Kakazu}}]{lai07}
{Lai}, K., {Huang}, J.-S., {Fazio}, G., {Cowie}, L.~L., {Hu}, E.~M., \&
  {Kakazu}, Y. 2007, ApJ, 655, 704

\bibitem[{{Maloney} \& {Black}(1988)}]{maloney88}
{Maloney}, P. \& {Black}, J.~H. 1988, ApJ, 325, 389

\bibitem[{{Rhoads} {et~al.}(2000){Rhoads}, {Malhotra}, {Dey}, {Stern},
  {Spinrad}, \& {Jannuzi}}]{rhoads00}
{Rhoads}, J.~E., {Malhotra}, S., {Dey}, A., {Stern}, D., {Spinrad}, H., \&
  {Jannuzi}, B.~T. 2000, ApJ, 545, L85

\bibitem[{{Riechers} {et~al.}(2006){Riechers}, {Walter}, {Carilli}, {Knudsen},
  {Lo}, {Benford}, {Staguhn}, {Hunter}, {Bertoldi}, {Henkel}, {Menten},
  {Weiss}, {Yun}, \& {Scoville}}]{riechers06}
{Riechers}, D.~A. et al. 2006, ApJ, 650, 604


\bibitem[{{Schaerer} \& {Pell{\'o}}(2005)}]{schaerer05}
{Schaerer}, D. \& {Pell{\'o}}, R. 2005, MNRAS, 362, 1054

\bibitem[{{Schinnerer} {et al.}(2008)}]{schinnerer08}
{Schinnerer}, E. {et al.} 2008, ApJ, 689, L5

\bibitem[{{Solomon} \& {Barrett}(1991)}]{solomon91}
{Solomon}, P.~M. \& {Barrett}, J.~W. 1991, in IAU Symposium, Vol. 146, Dynamics
  of Galaxies and Their Molecular Cloud Distributions, ed. F.~{Combes} \&
  F.~{Casoli}, 235

\bibitem[{{Solomon} \& {Vanden Bout}(2005)}]{solomon05}
{Solomon}, P.~M. \& {Vanden Bout}, P.~A. 2005, ARA\&A, 43, 677

\bibitem[{{Spergel} {et al.}(2007)}]{spergel07}
{Spergel}, D.~N. {et al.} 2007, ApJS, 170, 377

\bibitem[{{Stanway} {et~al.}(2004){Stanway}, {Bunker}, {McMahon}, {Ellis},
  {Treu}, \& {McCarthy}}]{stanway04}
{Stanway}, E.~R., {Bunker}, A.~J., {McMahon}, R.~G., {Ellis}, R.~S., {Treu},
  T., \& {McCarthy}, P.~J. 2004, ApJ, 607, 704

\bibitem[{{Taniguchi} {et al.}(2005)}]{taniguchi05}
{Taniguchi}, Y. {et al.} 2005, PASJ, 57, 165

\bibitem[{{Walter} \& {Carilli}(2008)}]{walter08}
{Walter}, F. \& {Carilli}, C. 2008, in Astronomical Society of the Pacific
  Conference Series, Vol. 395, Frontiers of Astrophysics: A Celebration of
  NRAO's 50th Anniversary, ed. A.~H. {Bridle}, J.~J. {Condon}, \& G.~C. {Hunt},
  49

\bibitem[{{Walter} {et al.}(2003)}]{walter03}
{Walter}, F. {et al.} 2003, Nature, 424, 406

\bibitem[{{Wei{\ss}} {et~al.}(2007){Wei{\ss}}, {Downes}, {Neri}, {Walter},
  {Henkel}, {Wilner}, {Wagg}, \& {Wiklind}}]{weiss07}
{Wei{\ss}}, A., {Downes}, D., {Neri}, R., {Walter}, F., {Henkel}, C., {Wilner},
  D.~J., {Wagg}, J., \& {Wiklind}, T. 2007, A\&A, 467, 955

\bibitem[{{Wei{\ss}} {et~al.}(2001){Wei{\ss}}, {Neininger}, {H{\"u}ttemeister},
  \& {Klein}}]{weiss01}
{Wei{\ss}}, A., {Neininger}, N., {H{\"u}ttemeister}, S., \& {Klein}, U. 2001,
  A\&A, 365, 571

\bibitem[{{Yan} \& {Windhorst}(2004)}]{yan04}
{Yan}, H. \& {Windhorst}, R.~A. 2004, ApJ, 600, L1

\end{thebibliography}

\end{document}